\documentclass[aps,prl,twocolumn,groupedaddress,showkeys, showpacs,nofootinbib]{revtex4-1}
\usepackage{graphicx}
\usepackage{color}
\usepackage{amsmath}

\newcommand{\vect}[1]{\boldsymbol{#1}}

\begin{document}

\title{
  From amorphous aggregates to polymer bundles:\\
  The role of stiffness on structural phases in polymer aggregation
}
\author{Johannes Zierenberg}
\email[]{zierenberg@itp.uni-leipzig.de}
\author{Wolfhard Janke}
\email[]{janke@itp.uni-leipzig.de}
\affiliation{Institut f\"ur Theoretische Physik, 
             Universit\"at Leipzig, 
             Postfach 100\,920, 
             D-04009 Leipzig, 
             Germany
           }
\date{\today}

\begin{abstract}
  We study the aggregation transition of a finite theta-polymer system in
  dependence on the bending stiffness $\kappa$ with the help of parallel
  multicanonical simulations. In order to distinguish amorphous aggregates from
  polymer bundles we introduce an order parameter, measuring the correlation of
  the end-to-end vectors. With the help of this order parameter, we construct
  generic $T$-$\kappa$ phase diagrams for systems with $2$ and $8$ polymers and
  discuss the occurring phases from amorphous aggregates to bundle structures.
  For an intermediate stiffness range we find multiple aggregated phases which
  change with increasing number of polymers and discuss their nature with the
  help of microcanonical analyses.  We show that the stiffness of semiflexible
  theta polymers is the distinguishing parameter for the emergent structural
  motifs.
\end{abstract}



\maketitle


Understanding the mechanism of polymer aggregation is of relevance for a wide
range of research, from biophysical actin networks to the design of materials
with certain mechanical properties.  Another important subject is protein
aggregation which is associated with several human diseases like Alzheimer's
disease, Parkinson's disease and diabetes II \cite{Chiti2006}.  In this context,
the distinction between amorphous aggregates and amyloid fibrils was argued to
depend on the free-energy barriers~\cite{Yoshimura2012}.  The concern of bundle
formation has also been addressed recently in the limit of rather stiff polymers
for actin networks. This includes the study of unbinding transitions for two and
more parallel filaments (modeling the polymers as wormlike
chains)~\cite{KierfeldBundles}, the twisting of filaments in a (wormlike) bundle
model ~\cite{FreyWLCBundles, GrasonBunldes,Turner2003}, and a theoretical
discussion of the influence of kinetics leading to the proposition of an
experiment to study this explicitly~\cite{Kraikivski2008}.  Since biopolymers
are rather stiff, the question arises as to which properties can be associated
to stiffness alone.

More generally, it is of great interest to unravel which properties can be
reproduced already with a simple, generic model of theta polymers relying merely
on excluded volume, short-range attraction and stiffness. Coarse-grained models
for short peptides were used to show that the aggregation of peptides is a
phase-separation process~\cite{JunghansPeptide}.  Similarly for exemplary
semiflexible polymers, it was shown that the aggregation transition may be
accompanied by an additional freezing transition depending on the
stiffness~\cite{JunghansPolymer}.  A single semiflexible polymer has been
studied using mean-field calculations~\cite{Orland1996} and PERM chain-growth
simulations of lattice models~\cite{Grassberger1997,Prellberg2010} finding
essentially an extended, a collapsed and a solid phase. Using a comparably short
off-lattice bead-spring polymer, it has only recently been shown that already
for a $30$mer the effect of stiffness leads to a phase diagram with a multitude
of conformational phases, ranging from globular to toroidal
structures~\cite{Seaton2013}.  Another generic model, the tubelike polymer,
shows a similar broad spectrum of complex
conformations~\cite{MaritanTubeLike,VogelTubeLike,MaritanTubeLikeSpecific,BanavarTubeLikeMany},
depending on its radius of curvature controlling the stiffness. With additional
specific interactions~\cite{MaritanTubeLikeSpecific}, the model has been
recently applied to protein aggregation~\cite{AuerAggregation}. 

In this study, we consider a semiflexible theta-polymer model which leads to a complex
interplay of collapse and aggregation with stiffness. We investigate the full
semiflexible range of polymer aggregation, from flexible to stiff. Addressing the
necessity to distinguish amorphous aggregates from polymer bundles, we introduce an order
parameter measuring the correlation between the polymer end-to-end vectors. 
Our results suggest that the polymer stiffness plays a key role in whether the
system forms an amorphous aggregate or a correlated polymer bundle.

The coarse graining of semiflexible polymers leads to a variety of models
ranging from lattice to continuum formulations.  We employed a common
bead-spring model with additional bending stiffness.  Here, the elastic bonds
are described by the finitely extensible nonlinear elastic (FENE) potential
\mbox{$V_{\rm FENE}(r)=-\frac{K}{2}R^2\ln\left(1-[(r-r_0)/R]^2\right)$}, where
we set $r_0=0.7$, $R=0.3$ and $K=40$ \mbox{following
\cite{Binder2001,Schnabel2009}}.  All other nonbonded monomer-monomer
interactions are modeled by the Lennard-Jones potential \mbox{$V_{\rm LJ}(r) = 4
\epsilon \left[ (\sigma/r)^{12} - (\sigma/r)^6 \right] =
\epsilon\left[(r_0/r)^{12}-2(r_0/r)^6\right]$} with $\epsilon=1$,
$\sigma=r_0/2^{1/6}$ and cutoff at $r_c=2.5\sigma$.  In order to avoid a jump in
the energy, the Lennard-Jones potential is shifted by $V_{\rm LJ}(r_c)$.  In
accordance with Ref.~\cite{JunghansPolymer}, there is no distinction between
intra-chain contacts and inter-chain contacts.  The Lennard-Jones interaction
accounts for self- and mutual-avoidance and sets the scale of short-range
attraction.  The bending stiffness is modeled by an energy contribution from the
discretized curvature of the polymers $E_{\rm bend} =
\kappa\sum(1-\cos\theta_i)$, where $\theta_i$ is the angle between neighboring
bonds.  We measure $\kappa$ and other energies in units of $\epsilon$ and
lengths in units of the bond length $r_0$. For typical interaction energies
($\epsilon\approx2\mathrm{kJ/mol}$~\cite{Martini2007}) room temperature
corresponds to $T\approx1$ in our dimensionless units (where $\epsilon = k_B =
1$).  The polymer system was simulated in a cubic box of extension $L$ with
periodic boundary conditions at fixed density $\rho=NM/L^3=0.001$, where $M$ is
the number of polymers of length $N$. In this study we considered $M=2,8$ and
$N=13$.

Despite the simplicity of this model, the extended parameter space requires the
use of state of the art simulation methods such as Wang-Landau
sampling~\cite{WangLandau} or the multicanonical method~\cite{MUCA, Janke1998}.
We applied a novel parallelization of the multicanonical
method~\cite{ZierenbergPMUCA} with up to $128$ cores, sampling a broad
temperature range for many $\kappa$ values in the desired region from flexible
to stiff polymers.  This will be achieved by introducing a weight function
$W(E)$ that is iteratively adapted so that updates ($\mu\rightarrow\mu'$) which
are accepted with probability $\min(1,W(E_{\mu'})/W(E_{\mu}))$ lead to a flat
energy distribution.  We considered a combined set of updates including simple
single-monomer shifts, bond rotations and polymer translation but also the more
sophisticated double bridging move, which proved to be important especially in
the amorphous region. In addition, we applied adapted variable update ranges
with bias correction~\cite{Schnabel2011Adv} to optimize the acceptance rates in
every part of the energy landscape.  The canonical and microcanonical averages
are obtained afterwards by standard reweighting techniques.

In order to distinguish the occurring phases, we considered the heat capacity
$C_V$, a ``phase'' separation parameter $\Gamma^2$ and our new order parameter.
The heat capacity is defined as the temperature derivative of the energy and
can be obtained from its thermal fluctuations $C_V=k_B\beta^2\left(\langle
E^2\rangle - \langle E \rangle^2\right)/NM$, with the inverse temperature
$\beta=(k_BT)^{-1}$.  The ``phase'' separation parameter is defined as radius
of gyration of the center of masses $\vect{r}_{\rm cm}^i$ of individual
polymers~\cite{JunghansPeptide,JunghansPolymer}, namely
$\Gamma^2=\frac{1}{2M^2}\sum_{i,j}\left(\vect{r}_{\rm cm}^i-\vect{r}_{\rm
cm}^j\right)^2$. This parameter will be small if all polymers are close
together in an aggregate; and large if separated in the
soluble phase. In addition, amorphous aggregates should be
distinguished from correlated structures such as bundles or ordered sheets.  It
can be observed that in the case of stiff polymer bundles, the end-to-end
vectors $\hat{\vect{R}}_i$ (normalized to unit length) of the individual
polymers are highly correlated, since the polymers align parallel. On the
other hand, when the polymers are separated or in the amorphous phase, the
end-to-end vectors are uncorrelated and their relative orientation is random.
This is why we introduce an end-to-end correlation parameter $C_R$ normalized
in such a way that for uncorrelated vectors the parameter assumes the value
$1/3$ while it tends to $1$ in the correlated case:
\begin{equation}\label{eq:orderParameter}
  C_R = \frac{2}{M(M-1)}\sum_{i<j} \left(\hat{\vect{R}}_i\cdot \hat{\vect{R}}_j\right)^2.
\end{equation}
Alternatively, one may consider the nematic order parameter~\cite{Allen1993} at
the additional cost of computing a full bond-bond interaction tensor.

\begin{figure}[t]
  \includegraphics{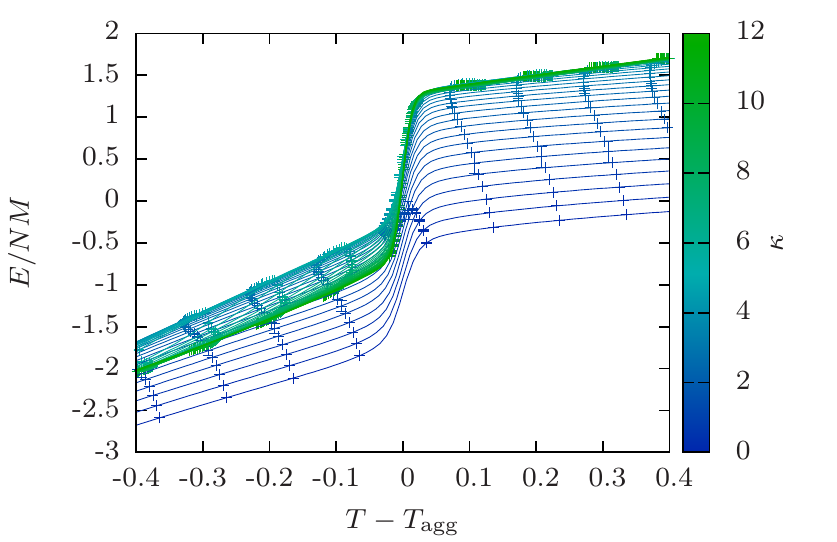}
  \includegraphics{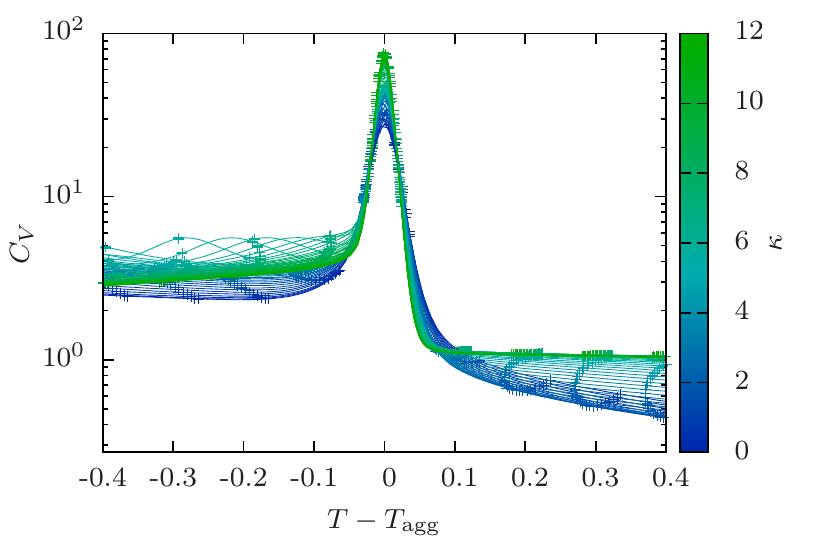}
  \vspace{-1.3em}
  \caption{(Color online)
    Effect of stiffness on the average energy and heat capacity in the
    canonical ensemble for 8 polymers of length $N=13$. The graphs are colored according to the bending
    stiffness $\kappa$ in the interval $[0,12]$ with $\Delta\kappa=0.2$. In all
    cases, the data display a sharp, discontinuous transition when the polymers
    aggregate.
    \label{fig:Canonical}
  }
\end{figure}

\begin{figure}[t!]
  \includegraphics{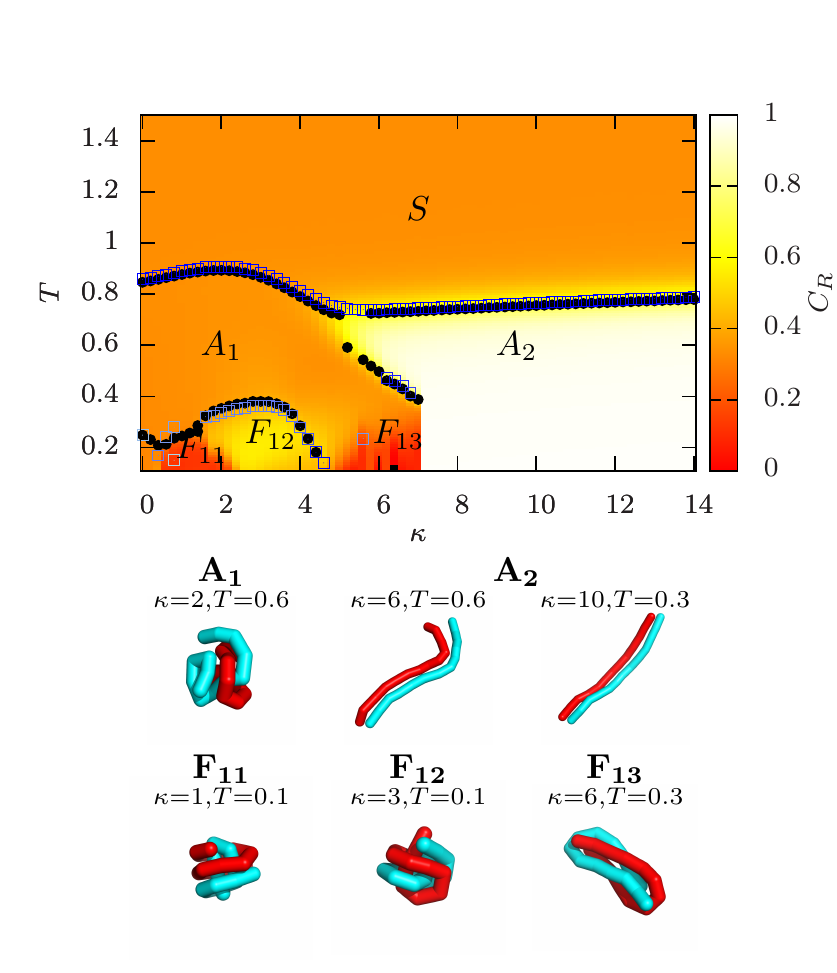}
  \vspace{-1.3em}
  \caption{(Color online)
    Phase diagram for 2 polymers consisting of \mbox{$N=13$} monomers each,
    combining the surface plot of the end-to-end correlation parameter $C_R$,
    the maxima of the heat capacity (black dots) and the temperature derivative
    of the phase separation parameter $\Gamma^2$ (blue squares).  We identify
    several structural phases, namely $S$ (soluble), $A$ (aggregated) and $F$
    (frozen), and present typical conformations for the low-temperature phases.
    \label{fig:phaseDiagram2}
  }
\end{figure}

Figure~\ref{fig:Canonical} shows the average energy and the heat capacity for a
system of 8 polymers in the canonical ensemble at various stiffnesses (encoded
in the line color). For clarity, we shifted the temperature axis to center
around $T_{\mathrm{agg}}$, defined as the temperature of the largest peak in
$C_V$. We obtained the error bars from time-series reweighting and the lines
with higher resolution from histogram reweighting.  The sharp drop in energy and
the peak in the heat capacity show that the aggregation transition of 8 short
semiflexible polymers is a discontinuous transition between an entropy
dominated, soluble regime at high $T$ and an energy dominated, aggregated regime
at low $T$. On closer inspection, we observe that the size of the energy jump
increases with stiffness as does the peak of the heat capacity.  Below the
aggregation transition, we observe less pronounced peaks indicating continuous
phase transformations which will be discussed below.  

A possible way to investigate the thermodynamics of large-scale systems is to
study the limit of an increasing number of polymer chains, considering the polymer length as
a fixed chemical property~\cite{Zierenberg2014JCP}. For a detailed $\kappa$
range, this is clearly unfeasible with the current computational resources.  A
finite-size scaling study would have to focus on a few $\kappa$ values that may
be chosen from the following diagrams. 

\begin{figure}[t!]
  \includegraphics{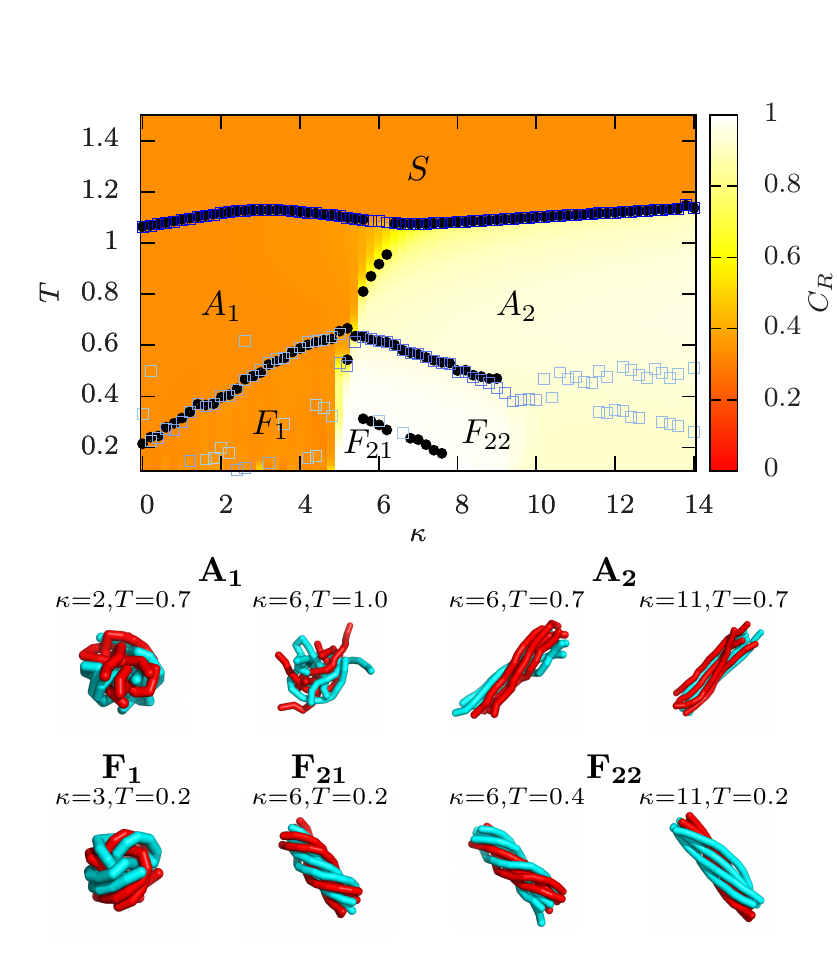}
  \vspace{-1.3em}
  \caption{(Color online)
    Same as Fig.~\ref{fig:phaseDiagram2} for $8$ polymers.
    \label{fig:phaseDiagram8}
  }
\end{figure}
Our main results are presented as ``phase''-diagrams in
Fig.~\ref{fig:phaseDiagram2} \mbox{($2\times13$mers)} and
Fig.~\ref{fig:phaseDiagram8} \mbox{($8\times13$mers)}. Since the systems are
finite, the (pseudo) transitions between structural phases should not be
confused with thermodynamic phase transitions (which would require taking the
thermodynamic limit). Instead, both structural phase diagrams give an overview
of what happens in the aggregation process of an entire class of semiflexible
polymers over a wide temperature and stiffness range.  We obtained the diagrams
from extensive parallel multicanonical simulations at $71$ fixed $\kappa$ values
in the interval $[0,14]$ with $\Delta\kappa=0.2$. This clearly covers the range
from flexible ($\kappa=0$) to rather stiff ($\kappa\approx N$) polymers. The
obtained picture is supported by exemplary simulations for longer and more
chains.  Even with our advanced techniques, the computational effort is still
demanding and cannot easily be increased. The phase diagrams show the landscape
of the end-to-end correlation parameter $C_R$, combined with transition points
determined from the peak locations of the heat capacity (black dots) and the
temperature derivative of the phase separation parameter (blue squares). In the
latter case, the squares are color coded from dark blue (rather strong signals)
to light blue (rather weak signals, which in some cases may be merely small
fluctuations without physical significance). In addition, we present typical
conformations from different regions for selected $\kappa$ and $T$. 

We distinguish between the fragmented or soluble phase ($S$), in which all
polymers are independent of each other; the aggregated phases ($A$), where all
polymers are located close to each other; and the ``frozen'' phases ($F$), which
occur when lowering the temperature even further. The term ``frozen'' refers to
rather well-ordered structures compared to fluctuating conformations in the $A$
phases. At the aggregation transition, we observe the peaks in the heat capacity
and the temperature derivative of the phase separation parameter close to each
other, this being typical for first-order like transitions. In both phase
diagrams we can clearly see that rather flexible polymers \mbox{(low $\kappa$)}
aggregate into a phase $A_1$ where $C_R$ is \mbox{roughly $1/3$}, while for
stiff polymers the aggregated phase $A_2$ is described by $C_R\approx1$. Looking
at typical conformations we observe amorphous structures for aggregates of
rather \mbox{flexible polymers ($A_1$)} and stretched bundles for \mbox{stiffer
polymers ($A_2$)}. Going to even lower temperatures, stiff polymers rearrange
within the bundle phase whereas more flexible polymers show an additional
``freezing'' transition similar to the single polymer
case~\cite{MaritanTubeLike, VogelTubeLike, Schnabel2009, Seaton2013} but with
the additional possibility to wrap around each other.

Following the characterization of the aggregated phases via the end-to-end
correlation, the ``frozen'' phases are labeled $F_{1i}$ for different subphases
showing uncorrelated conformations and $F_{2i}$ for distinguishable subphases
that show bundle like characteristics.  In the latter case, typical
conformations of this correlated aggregate may be described as twisted rods,
recapturing the qualitative behavior of the wormlike bundle
model~\cite{FreyWLCBundles} for sufficiently large stiffness at fixed $N$. Using
a twisted-bundle parameterization, we verified that at zero temperature the
energy minimum is lower than that for parallel rods. This can be explained by
the large energetic gain from maximizing Lennard-Jones contacts when parallel
bundles are twisted compared to the energetic loss from bending. This is in
agreement with analytic considerations of twisted fibers~\cite{Turner2003} and
numeric studies on several tubelike polymers~\cite{BanavarTubeLikeMany}. In this
regime of rather stiff theta polymers, the (discrete) wormlike chain is a good
approximation and we may express the persistence length to leading-order as
$l_p/r_0\approx\kappa/k_BT$.\footnote{
  Note that for rather flexible theta polymers, there has been a recent debate
  on the definition of the persistence length~\cite{BinderLp}.
} 
Hence, within the bundle phase of our diagrams the
length ratio $l_p/Nr_0$ is of order unity. This matches typical scales of
amyloid fibrils, where the pitch length is of the order of the protofibril
length (which, however, are both much larger)~\cite{Serpell2000}.  Additional
specific interactions such as hydrogen bonds and hydrophobic effects may
stabilize or destabilize structural motifs.  This has been shown for single
proteins within the tubelike model~\cite{MaritanTubeLikeSpecific}; and applied
to peptide aggregation it was possible to reproduce stable fibrillar
structures~\cite{AuerAggregation}.

There is an intermediate regime \mbox{$\kappa\approx4 \dots 8$} for the two system sizes,
where an additional transition below the aggregation temperature differs noticeably from
$2$ to $8$ polymers.  By lowering the temperature, $2$ polymers in this region show an
initial correlation followed by a decorrelation of the end-to-end vectors.  While typical
conformations in the initial correlated phase (Fig.~\ref{fig:phaseDiagram2}, $A_2$) could
be described as elongated aggregates, the conformations in the uncorrelated phase
(Fig.~\ref{fig:phaseDiagram2}, $F_{13}$) show a different structure that reminds of
entangled hairpins. These structures occasionally occur in the initial aggregated phase
but become more probable at lower temperatures. The entangled hairpins may have a slight
twist and in some cases the U-like polymer structures entangle perpendicularly
($C_R\approx 0$). For $4$ polymers in this regime (not shown here) we observe a similar
behavior aggregating first into a correlated phase followed by a decorrelated phase, while
$8$ polymers demonstrate the reverse situation. For example along $\kappa=6$, with
decreasing temperature the aggregation transition first results in an uncorrelated, or
amorphous-like, \mbox{aggregate (Fig.~\ref{fig:phaseDiagram8}, $A_1$)}. Lowering the
temperature further, the end-to-end vectors start to correlate and the polymers form
bundles (Fig.~\ref{fig:phaseDiagram8}, $A_2$). At even lower temperatures, the polymer
bundles undergo an additional structural transition into the frozen phases $F_{22}$ and
$F_{21}$. Notice that varying $\kappa$ at fixed low $T$ leads to a narrow crossover from
amorphous aggregates to polymer bundles at about $\kappa\approx 7$ for $M=2$ and
$\kappa\approx 5$ for $M=8$.


In order to investigate the difference between $2$ and $8$ interacting polymers
in more detail, we employed a microcanonical analysis~\cite{Microcanonical,
MicroJanke}, which was proven to be particularly suitable for the study of
structural phases in finite
systems~\cite{JunghansPeptide,JunghansPolymer,Schnabel2011}.  In order to
achieve this, we took advantage of the employed generalized ensemble simulations
which enable us to calculate the microcanonical entropy $S(E)$, up to an
additive constant, together with its first and second derivatives.  The first
derivative yields the microcanonical caloric inverse temperature
$\beta(E)=\frac{\partial S}{\partial E}$ which encodes first- and second-order
transitions in its inflection points.  These points are seen in the second
derivative $\gamma(E)=\frac{\partial^2 S}{\partial E^2}$ as maxima with
$\gamma>0$ for first-order transitions and $\gamma<0$ for second-order
transitions~\cite{Schnabel2011}.  Note that this is just the inverse
microcanonical specific heat, \mbox{$\gamma(E)= -\left[C_{\rm
micro}(E)/\beta(E)^2\right]^{-1}$}.  Since the canonical energy is increasing
with temperature, transitions at lower energies may be associated with
transitions at lower temperatures.

\begin{figure}[t]
  \includegraphics{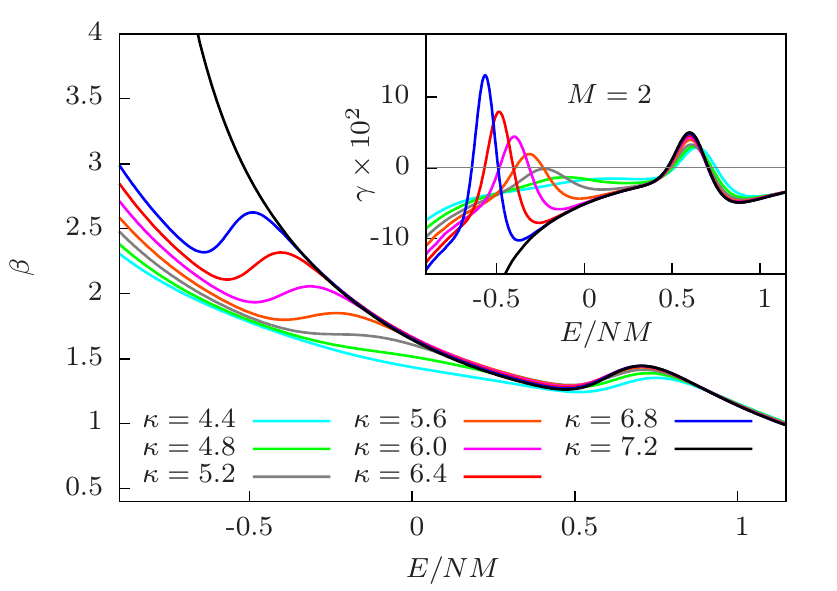}
  \includegraphics{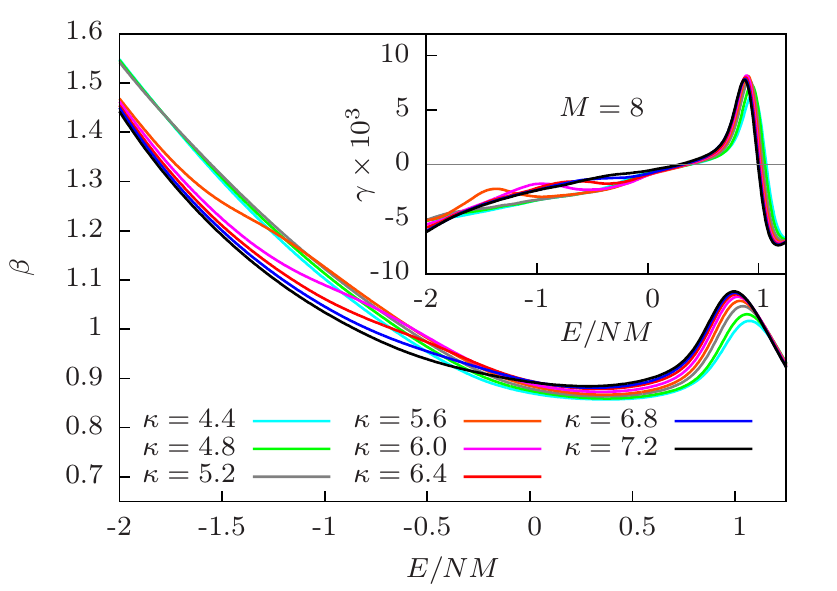}
  \caption{(Color online)
    Microcanonical analysis: The first ($\beta$) and second ($\gamma$) derivatives of 
    the microcanonical entropy for $M=2$ (top) and $8$ (bottom) polymers show the
    occurrence of an additional transition for the intermediate stiffness regime 
    besides the first-order like aggregation transition ($\gamma>0$ peak).
    For 2 polymers an additional first-order transition ($A_2\rightarrow F_{13}$) occurs.
    For 8 polymers the additional transition ($A_1\rightarrow A_2$) is of second order
    ($\gamma<0$ peak).
    \label{fig:MicroCanonical}
  }
\end{figure}

\mbox{Figure~\ref{fig:MicroCanonical}} shows $\beta(E)$ and $\gamma(E)$ for
several $\kappa$ values in the intermediate regime for $2$ and $8$ polymers.  In
both cases, we can see a first-order signature ($\gamma>0$ peak) for the
aggregation transition at larger energies. Also, an additional peak appears at
lower energies for the intermediate $\kappa$ regime from $\kappa\approx5$ to
$\kappa\approx7$ consistent with the observation from the canonical picture,
which vanishes for smaller or larger $\kappa$. This transition shows a clear
difference between the case of $2$ and $8$ polymers.  For 2 polymers and low
$\kappa$ the peak at smaller energies is weak and below zero, suggesting a
second-order transition, while for larger $\kappa$ the peak becomes pronounced
and larger than zero indicating a first-order transition from correlated
polymers directly into the ``frozen`` entangled-hairpin phase ($A_2\rightarrow
F_{13}$).  In the case of 8 polymers the (less pronounced) peak at lower
energies is below zero indicating that the corresponding transition from
amorphous aggregates to polymer bundles ($A_1\rightarrow A_2$) is of second
order.
  
\begin{figure}[t]
  \includegraphics{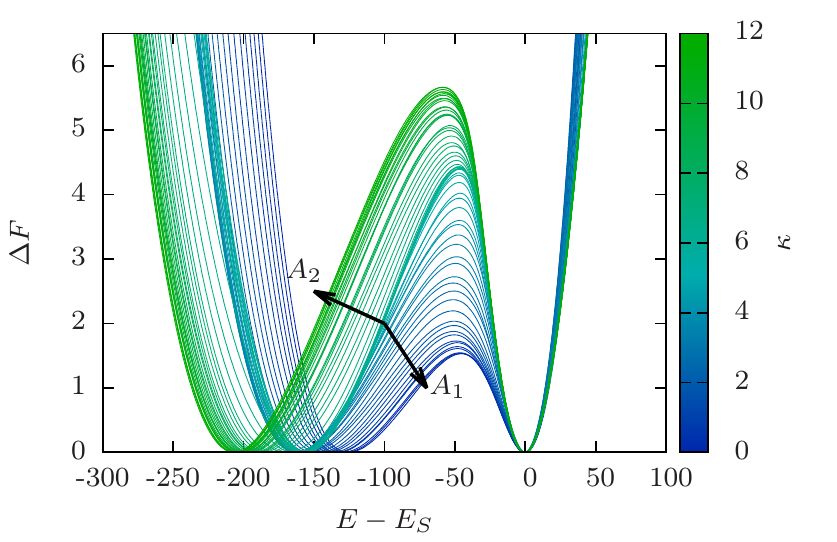}
  \caption{(Color online)   
    Free-energy barrier at the equal-height aggregation temperature depending on
    the stiffness $\kappa$, color coded as in Fig.~\ref{fig:Canonical}. $E_S$ is
    the energy of the soluble phase.
    \label{fig:FreeEnergy}
  }
\end{figure}

Finally, having settled the first-order nature of the aggregation transition, we
address the question of the free-energy barrier accompanying such a transition.
Figure~\ref{fig:FreeEnergy} shows an example of the 8 polymer system. The free
energy is obtained from the energy probability distribution at equal-height
temperature $T_{\rm eqh}$, namely $F_{\rm eqh}(E)=-k_B T_{\rm
eqh}\ln\left(P_{\rm eqh}(E)\right)$ and $\Delta F = F_{\rm eqh}(E)-F_{\rm min}$.
The minima of this free energy correspond to the equilibrium phases at
coexistence, the soluble ($S$) phase at $E_S$ and the aggregated ($A$) phase at
$E_A$. Already for this finite system size, the existence of a local maximum, or
barrier, between these two phases reconfirms the first-order nature. The barrier
in the free energy clearly depends monotonically on the stiffness. This supports
the claim that the free-energy barrier for amorphous aggregation is lower than
for aggregation into bundles~\cite{Yoshimura2012}.  In addition, one may
qualitatively distinguish the amorphous regime $A_1$ (blue) from the bundle
regime $A_2$ (green), indicated by the two arrows in Fig.~\ref{fig:FreeEnergy}.

In summary, employing sophisticated simulation and analysis methods rooted in a
generalized ensemble approach, we have been able to map out the generic
$T$-$\kappa$ ``phase'' diagram, from flexible to stiff polymer systems with 2
and 8 short polymer chains. The thus obtained systematic overview emphasises the
key role of stiffness in polymer aggregation, leading from amorphous aggregates
to correlated structures such as polymer bundles.  In the intermediate stiffness
regime, we identified two aggregated phases at constant stiffness, separated
(for larger systems) by a second-order like transition as indicated by the
microcanonical analysis.  For the first-order aggregation transition of the 8
polymer system, we provided strong evidence that the free-energy barrier for the
transition to polymer bundles is indeed higher than for the transition into
amorphous aggregates, as discussed recently in a slightly different context.
Our systematic study supports the claim that the combination of excluded volume
and short-range attraction with stiffness is the basic mechanism for the
formation of this variety of structural motifs.  Similar motifs occur, e.g.,
after bundling into mature amyloid fibrils~\cite{Talaga2008} and within
viruses~\cite{Noda2006}. In the context of material design, it has been observed
for patchy particles~\cite{Glotzer2004} and upon adsorption to
nano-wires~\cite{Vogel2004}.  In general, such a behavior is quite generic for
biopolymers, which typically show rather stiff characteristics.

\begin{acknowledgments}
We would like to thank Thierry Platini for helpful discussions on the introduced
order parameter and Martin Marenz for the joint development of a Monte Carlo Simulation Framework.
The computing time provided by the John von Neumann Institute for Computing (NIC) under grant No.~HLZ21
on the supercomputer JUROPA at J\"ulich Supercomputing Centre (JSC) is gratefully acknowledged.
This work has been partially supported by
 the Leipzig Graduate School of Excellence GSC185 ``BuildMoNa'',
 the Collaborative Research Center SFB/TRR 102 (project B04) and
 the Deutsch-Franz\"osische Hochschule (DFH-UFA) under grant No.\ CDFA-02-07.
The project was funded by the European Union and the Free State of Saxony.
\end{acknowledgments}

\bibliographystyle{model1-num-names}

\end{document}